\documentclass[prb,aps,floats,amssymb,showkeys,showpacs,preprint,
 bibnotes, tightenlines]{revtex4}
\usepackage{graphicx}
\usepackage{graphics}
\usepackage{epsfig,bm}
\usepackage{rotating}
\begin{document}

\title{Yang-Lee edge singularities from extended  activity \\
 expansions of the dimer density  for bipartite lattices \\
 of dimensionality $2 \le d \le 7$ \\ }

\author{P. Butera}

\email{paolo.butera@mib.infn.it}

\affiliation
{Dipartimento di Fisica Universita' di Milano-Bicocca\\
and\\
Istituto Nazionale di Fisica Nucleare \\
Sezione di Milano-Bicocca\\
 3 Piazza della Scienza, 20126 Milano, Italy}

\author{M. Pernici} 

\email{mario.pernici@mi.infn.it}

\affiliation
{Istituto Nazionale di Fisica Nucleare \\
Sezione di Milano\\
 16 Via Celoria, 20133 Milano, Italy}

\date{\today}

\date{\today}
\begin{abstract}
We have extended, in most cases through 24th order, the series
 expansions of the dimer density in powers of the activity in the case
 of bipartite ((hyper)-simple-cubic and (hyper)-body-centered-cubic)
 lattices of dimensionalities $2 \le d \le 7$. A numerical analysis
 of these data yields estimates of the exponents characterizing the
 Yang-Lee edge-singularities for lattice ferromagnetic spin-models as
 $d$ varies between the lower and the upper critical
 dimensionalities. Our results are consistent with, but more extensive
 and sometimes more accurate than those obtained from the existing
 dimer series or from the estimates of related exponents for lattice
 animals, branched polymers and fluids. We mention also that it is
 possible to obtain estimates of the dimer constants from our series
 for the various lattices.

\end{abstract}
\maketitle

\section{Introduction}
Let us consider a classical spin-1/2 Ising system with
nearest-neighbor ferromagnetic interactions, in the presence of an
external magnetic field $H$, on a finite $d$-dimensional lattice of
$N$ sites.  The partition function can be written as
\begin{equation}
Z_N(H,T)=\sum_{conf} 
{\rm exp}\Big [ K \sum_{<ij>} s_is_j +h \sum_i s_i \Big ]. 
\label{Isipf}
\end{equation}
Here $s_i= \pm 1$ denotes an Ising spin variable associated to the
 site $i$ of the lattice. The first sum extends to all configurations
 of the spins, the second to all distinct pairs $<ij>$ of
 nearest-neighbor spins, the third to all spins. We have set $K=
 J/k_BT$, with $T$ the temperature, $J>0$ the exchange coupling of the
 spins, $k_B$ the Boltzmann constant, and $h=H/k_BT$ the reduced
 magnetic field.  It is a classical result\cite{YL1,YL2} that, for
 real $T$, $Z_N(H,T)$ has a sequence of zeroes that can occur only on
 the imaginary axis of the complex $h=h'+ih''$ plane. This statement
 remains true for a wide variety of models with ferromagnetic
 couplings, including Ising models of arbitrary spin\cite{grif},
 $n$-vector models\cite{dunlop,new,lieb} (at least for $n<4$), etc..
 As long as the number $N$ of sites remains finite, these zeroes give
 rise to a sequence of logarithmic branch-points of the free energy.
 For $T$ above its critical value $T_c$, 
when the thermodynamic limit $N \rightarrow \infty$ is taken, these
 cuts coalesce into two continuous lines of singularities along the
 imaginary axis of the $h$-plane that extend to infinity and originate
 at the ``Yang-Lee(YL) edges'', $h''=\pm h''_{YL}(T)$, with
 $h''_{YL}(T)>0$.  The
 partition function must then be nonzero and the free energy must be
 analytic in a strip containing the real axis of the complex reduced
 field plane, so that no phase transition can occur for $T >T_c$.  As
 $T \rightarrow T_c$, we have $h_{YL}''(T) \rightarrow 0$, so the gap
 between the endpoints of the singularity lines shrinks and they tend
 to pinch the real axis of the complex field plane giving rise, for
 $T<T_c$, to the characteristic discontinuity associated with the
 spontaneous magnetization.

 At  fixed $T> T_c$, in the vicinity of a YL edge, the magnetization per spin
is expected to exhibit a power-law behavior\cite{kort}
\begin{equation} 
M(H,T) \sim |h \pm
ih_{YL}''(T)|^{\sigma}
\label{magn}
\end{equation}
 controlled by an exponent $\sigma$. 

The main features of these YL edge-singularities have been extensively
studied. {\it A priori} the exponent $\sigma$ might depend on the
temperature and the particular features of the model, however it was
observed that it is universal, i.e. it depends only on the spatial
dimensionality, but not on the structure of the lattice on which the
spin model is defined.  Moreover, due to the presence of the magnetic
field, this exponent does not depend on the number of components of
the spin. Hence for an $n$-vector ferromagnetic system, it must be the
same as for an Ising system.  Finally, the study of solvable models in
one dimension or for $d\rightarrow \infty$ (the mean-field model) or,
for any dimension, in the $n \rightarrow \infty$ limit (the spherical
model), and finally the numerical study\cite{kort} of non-trivial
Ising systems in $d=2$ and $d=3$ dimensions, indicates that, for $T
>T_c$, the exponents $\sigma=\sigma(d)$ are temperature-independent.
Due to these universality features, the properties of the YL edges
have been related\cite{fisher} to those of the ordinary critical
points in the theory of second-order phase transitions.  In this respect, 
an important
step forward was to show that a renormalization-group approach to the
calculation of the exponent $\sigma$ can be formulated\cite{fisher} in
terms of an effective Landau-Ginzburg one-component scalar field
theory with a cubic interaction and a purely imaginary coupling. This
entails the upper critical dimensionality $d_c=6$, at and above which
the exponent takes its mean-field value $\sigma=1/2$. Moreover
$\sigma$ can be expressed as an expansion in powers of $\epsilon=6-d$, 
whose coefficients have been so far computed\cite{fisher,bon} through
the third order.  See Eq.(\ref{sigmaeps}) below.

The interest of an accurate determination of $\sigma(d)$ is enhanced
by the realization that the YL edge exponent is also related to other
exponents characterizing the behavior of several quite different
systems:

a) The pressure for $d$-dimensional fluids, with repulsive-core
 interactions among their constituents, exhibits an unphysical
 singularity located at negative values of the activity and showing
 universal exponents\cite{poland,baram,lai,park}
 $\phi(d)=\sigma(d)+1$. A similar property is true for lattice systems
 of hard squares (and higher-dimensional solids).

b) The number-per-site of directed branched polymers with $s$ bonds
(or of directed site- or bond-animals\cite{stauf,gutlib} of size $s$)
on a $d$-dimensional lattice, has a scaling behavior characterized by
a power of $s$ with universal exponent\cite{card,stan,breuer}
$\phi_D(d)=\sigma(d-1)+1$.  The scaling behavior of the directed
branched polymers has been mapped\cite{imbrie} into that of the
aforementioned fluids at the unphysical singularity, in one fewer
space dimension.

c) The universal exponents characterizing the behavior of the
number-per-site of large isotropic branched polymers\cite{par} in a
good solvent (or, equivalently\cite{lub}, of undirected site- or
bond-animals) on a $d$-dimensional lattice can be expressed as
$\phi_I(d)=\sigma(d-2)+2$. The scaling behavior of the isotropically
branched polymers has been mapped\cite{bryd} into that, mentioned
above, of the fluids in two fewer dimensions.

 d) A field theoretic model\cite{luben} of the Anderson localization
exhibits the same critical behavior as the isotropic branched polymers
 mentioned  above in c).

 Refs.[\onlinecite{lai,adler}] provide us with tables of numerical
 estimates of these exponents for the systems of  a), b) and c) in
 spatial dimensionalities $0 \le d \le 8$, which corroborate these
 exponent identifications.  In particular, estimates of the exponent
 $\phi_I(d)$ for high-dimensional undirected site-animals have been
 obtained from 15th-order activity expansions\cite{adler}, and, more
 recently, have been complemented by extensive MonteCarlo
 simulations\cite{hsu} and exact enumerations\cite{luth}.

We have recently calculated\cite{bp1,bp2}, in most cases through 24th
 order, the high-temperature(HT) and low-field expansion of the
 magnetization in an external field for several models in the Ising
 universality class, including the conventional Ising spin-1/2 model,
 on a sequence of bipartite lattices of spatial dimensionalities $2
 \le d \le 7$. In this paper, we shall use these data, most not yet
 existing in the literature, to improve the accuracy and extend the
 range of direct estimates of the exponents of the YL
 edge-singularity.  Our work is based on the
 observation\cite{gaunt,baker,fisher,kurtze} that this task can be
 accomplished restricting to the study of the Ising magnetization in a
 particular HT limit, since the exponents $\sigma(d)$ are
 temperature-independent.  It was shown that, in this limit, for each
 value of $d$, the bivariate expansion of the magnetization in powers
 of the inverse temperature at low field reduces to a simpler
 univariate expansion of a quantity related to the density of a dimer
 system on the same lattice, in powers of a variable $z$ that can be
 viewed as the activity of the dimers. Hence, the estimates of the
 exponents $\sigma(d)$ can be obtained from the simpler study of a
 singularity in the complex $z$-plane.

 The paper is organized as follows. In the second Section, we describe
 briefly the dimer model and sketch the arguments relating the Ising
 magnetization and the dimer density.  In the third Section, we
 discuss the numerical estimates of the exponents $\sigma(d)$. In the
 fourth Section, we briefly mention the possibility of obtaining
 accurate estimates of the dimer constants from the dimer series for
 several lattices.  In the final Section, we summarize our work and
 draw some conclusions.
\begin{table}[ht]\scriptsize
\caption{ Expansion coefficients $d_n$ of the dimer density 
 per site $\rho=\sum_{n=1}^{\infty} d_n z^n$
 for various lattices.
 In the case of the square lattice, the  coefficients were already 
 known\protect\cite{kurtze} 
through order 17, in the simple-cubic lattice
 case through order 15, in the body-centered-cubic case through order 13. }

\begin{tabular}{|c|c|c|c|}
 \hline
          & sq  & sc& bcc\\
 \hline
$d_1$&    2&          3&            4\\
$d_ 2$&   -14&        -33&          -60\\
$d_ 3$&   116&        438&         1096\\
$d_ 4$& -1042&      -6381&       -22076\\
$d_5$&  9812&      98298&       471384\\
 $d_6$&-95288&   -1571646&    -10462752\\
 $d_7$&945688&   25804572&    238712352\\
 $d_8$&  -9537906& -432195261&  -5559491148\\
 $d_9$&  97398764& 7351521882& 131557495336\\
$d_{10}$&      -1004479624&  -126601633818&   -3152926387520\\
$d_{11}$&      10442811216&  2202345302028&   76350685086240\\
$d_{12}$&    -109291830952&-38634960958878&-1864887612147680\\
$d_{13}$&    1150263509280&682589371293612&45882795957148336\\
$d_{14}$&  -12164408791920&    -12133302712160964&    -1135919635021757184\\
$d_{15}$&  129177146454536&    216812614019536368&    28273604715543144816\\
$d_{16}$&-1376741271026898&  -3892138971898300893&  -707057878652500074156\\
$d_{17}$&14719835348283940&  70154203323444808578& 17755217609022707289048\\
$d_{18}$&-157827022198103624& -1269064065215177023170& 
-447499139067294036693120\\
$d_{19}$&1696509872615736256&   23030962300401009234828&
  11315759542146218070302592\\
$d_{20}$&-18277500804075889672& -419178020538794595807786
&-286983128267601041433972576\\
$d_{21}$&197318358676755340504& 7649309116068955304095452
&7297713321539491454122083408\\
$d_{22}$&-2134157389758234716560& -139919791900565246008916796
&-186023844148307269386048046560\\
$d_{23}$&23121845343883130936248&2564949986696274806668222464
&4752360937807056276521029006800            \\
$d_{24}$&-250895842743288634987656&-47113178583259985664916986702 
&-121654579352079670288510647161952 \\

 \hline
\colrule  
\end{tabular} 
\label{tab1}
\end{table}

\begin{table}[ht]\scriptsize
\caption{Expansion coefficients $d_n$   of the dimer density per site
$\rho=\sum_{n=1}^{\infty} d_n z^n$, in the case of 
 the h4sc and of the h4bcc lattices, for which no data exist in  
the literature. }
\begin{tabular}{|c|c|c|}
 \hline
           &{\rm h4sc}  & {\rm h4bcc}\\
 \hline

$d_{1}$ &           4&                     8\\
$d_{ 2}$&         -60&                  -248\\
$d_{ 3}$&        1096&                  9488\\
$d_{ 4}$&      -22100&               -403528\\
$d_{ 5}$&      473064&              18295568\\
$d_{ 6}$&   -10540512&            -865807424\\
$d_{ 7}$&   241719216&           42249956512\\
$d_{ 8}$& -5664784788&        -2109836426376\\
$d_{ 9}$&135024044344&       107268694276016\\
$d_{10}$&         -3262426210400&     -5532784189984768\\
$d_{11}$&         79709668733952&    288753867270427776\\
$d_{12}$&      -1965728629262720& -15218580836786546560\\
$d_{13}$&      48861010499029408&         808774465817914713216\\
$d_{14}$&-1222763683250863968&      -43288481720615472507456\\
$d_{15}$&30780446257662843696&     2331290979582407692061408\\
$d_{16}$& -778830611224069318356&  -126230584983092928283366536\\
$d_{17}$&19796376474243625682760& 6867529887763685727515233744\\
$d_{18}$&    -505225273989977429147424&  -375209852952369261768265528064\\
$d_{19}$&   12940687836519051181004448& 20577462138006148093915106417728\\
$d_{20}$& -332541786326958137163996960&  -1132370504978616213744537711827968\\
$d_{21}$& 8570723905802396225090813040& 62506221601565914680638405721677920\\
$d_{22}$&-221490566192354289867699314976&
-3459983056502070079122589397022126400\\
$d_{23}$& 5737972822751243066305678738704  
&192014799892871528814114123763838125600\\
$d_{24}$&-148983747147193194624766250874624 
&-10681019026724387660511023640151016838528\\
 \hline
\colrule  
\end{tabular} 
\label{tab2}
\end{table}

\begin{table}[ht]\scriptsize
\caption{ Expansion coefficients $d_n$ of the dimer density per site
$\rho=\sum_{n=1}^{\infty} d_n z^n$,   in the case of 
 the h5sc and of the h5bcc lattices, for which no data exist in  
the literature. }
\begin{tabular}{|c|c|c|}
 \hline
           &{\rm h5sc}  & {\rm h5bcc}\\
 \hline
$d_{1}$&                     5&           16\\
 $d_{2}$&          -95&        -1008\\
$d_{ 3}$&         2210&        78880\\
 $d_{4}$&       -56935&     -6888560\\
 $d_{5}$&      1560470&    643011936\\
$d_{ 6}$&    -44589470& -62772421632\\
$d_{ 7}$&   1312933060&6328669301376\\
$d_{ 8}$& -39545139015&   -653740982885040\\
$d_{ 9}$&1212383059070&  68823661085963680\\
$d_{10}$&    -37702721543530&-7356685696560126848\\
$d_{11}$&   1186280544459180&     796252824523662796416\\
$d_{12}$& -37692235412554270&  -87086303698430880977408\\
$d_{13}$&1207600893566962380& 9609234910400277365928640\\
$d_{14}$&     -38966834192905441260&     -1068376395044310231082721280\\
$d_{15}$&    1265202803884630177880&    119569512978084142014703005120\\
$d_{16}$& -41303509002873877637895& -13459302500635793010903317240880\\
$d_{17}$&1354882911889343633812990&1522782576346629681778159256296800\\
$d_{18}$&-44635036629478941618353330&-173070180804086640006367948602645120\\
$d_{19}$&1476101842809890620493385820&19750115971287989745172871861397745920\\
$d_{20}$&-48984496236897026046190228810&
 -2262063795052955748647989801450884168960\\
$d_{21}$& 1630653500450336878870631247220&
 259942605480019278589477461871133106000960\\
$d_{22}$&-54438143496599430247473673703380  
& -29961083663149896944399970822493200812776320\\
$d_{23}$&            &3462835569237061144802332732162000087319991360\\
$d_{24}$&            & -401237452358873371565154489237992019311699297280\\

 \hline
\colrule  
\end{tabular} 
\label{tab3}
\end{table}

\begin{table}[ht]\scriptsize
\caption{ Expansion coefficients $d_n$ of the dimer density per site 
$\rho=\sum_{n=1}^{\infty} d_n z^n$,  in the case of 
 the h6sc and of the h6bcc lattices, for which no data exist in  
the literature. }
\begin{tabular}{|c|c|c|}
 \hline
          &{\rm h6sc}  & {\rm h6bcc}\\
 \hline
$d_{1}$&    6  &    32\\
$d_{2}$&   -138  &      -4064\\
 $d_{3}$&   3900 &   643136\\
 $d_{4}$&   -122310  &   -113792032\\
$d_{5}$&   4086396           &  21547360832\\
 $d_{6}$& -142476792         &  -4271044245248\\
 $d_{7}$&5122735176  &  874932782274688\\
 $d_{8}$&-188517518310      & -183742680108707616\\
 $d_{9}$&7064821951140     &  39344571392293056704\\
$d_{10}$& -268661239644168  &   -8557358633880518454784\\
$d_{11}$&  10340279507701776    &    1885218050259702418922496\\
$d_{12}$& -402003973427568648&    -419794129141493673828232192\\
$d_{13}$& 15763048131921626880  & 94331821844075469462840829440\\
$d_{14}$&  -622646937215063631120 &-21363417876150273316847202904320\\
$d_{15}$&24752424038720810997240   &   
               4871104500436454268454972630464896\\
$d_{16}$&   -989526434040322269780870
&    -1117286822866349754715547920533865248 \\
$d_{17}$&  39754803422947748576043180    
&    257621581517072116407677681638058301760\\
$d_{18}$& -1604239977818820429337772040   
&  -59680091386808879450249399896382705283584\\
$d_{19}$&64993232282038522127491654560  
&  13883370669662028006386373064762884655595776 \\
$d_{20}$&-2642511808864143082376138558280  
& -3241891787443021814766940463936149652806538752 \\
$d_{21}$&107788019222920474225378253424840   
&759601317388101093048127502735990937113754258304\\
$d_{22}$& 
&-178534783824842834362110547423593615649270807425280 \\
$d_{23}$&
& 42081725576331603416392306775511862130364642844236928  \\
$d_{24}$& 
&-9944795581281129212782166232032493020106438172525218816  \\

 \hline
\colrule  
\end{tabular} 
\label{tab4}
\end{table}

\begin{table}[ht]\scriptsize
\caption{ Expansion coefficients $d_n$ of the dimer density per site 
$\rho=\sum_{n=1}^{\infty} d_n z^n$,  in the case of 
 the seven-dimensional sc and bcc lattices, for which no data exist in  
the literature. }
\begin{tabular}{|c|c|c|}
 \hline
          &{\rm h7sc}  & {\rm h7bcc}\\
 \hline

$d_{ 1}$&         7&        64\\
$d_{ 2}$&      -189&    -16320\\
$d_{ 3}$&      6286&   5193856\\
$d_{ 4}$&   -232337&   -1849727936\\
$d_{ 5}$&   9156882&  705432330624\\
$d_{ 6}$&-376866798&-281737586952192\\
$d_{ 7}$&   16002790188&  116325399366292992\\
$d_{ 8}$& -695763137601& -49250328402750744768\\
$d_{ 9}$&30814814722594&   21265341161536087474816\\
$d_{10}$&   -1385226427142474&-9328024156759223705208320\\
$d_{11}$&   63037109428097196& 4145112005630262044731153920\\
$d_{12}$&-2898142120281195662& -1862040622714788094682263408640\\
$d_{13}$& 134406753861096711820& 844180535052646908871372654226176\\
$d_{14}$&   -6280176867758451284340&-385755924146309028765643631630290940\\
$d_{15}$&  295357093420022257715616&177488316175328670895488110020974226176 \\
$d_{16}$&-13970161606657482586444833&
-82155985773613044917541164342517378676416 \\
$d_{17}$&664121988815559152932012890&
38231156319654300832459678139949723942242688 \\
$d_{18}$&   -31713748014463596772472407794&
-17875202142357193337368185100702670312872235520 \\
$d_{19}$&  1520542773828098297725351871436
&8393167978080707171882156986310088446859172518912\\
$d_{20}$&-73169317004302689750290322402042&
 -3956035371299664049518526770009878297187847196261376\\
$d_{21}$& &1871101505460166042039514382404078383796375971968146688 \\
$d_{22}$& &-887773576351504314786611216702175344355035369163473610240 \\
$d_{23}$&& 422432182192613348318985610505592297395940532680091395006720\\
$d_{24}$& &-201538729427533378265902478728701415730702344267711209153902592\\

 \hline
\colrule  
\end{tabular} 
\label{tab5}
\end{table}

\section{ Dimer density and Ising  magnetization}

The hard dimer model in two space dimensions was
 introduced\cite{Fowler} long ago to study the adsorption of diatomic
 molecules on a regular surface.  On a $d$-dimensional lattice of $N$
 sites it describes an assembly of linear objects each one covering a
 single bond of the lattice including the two endpoint sites, under
 the constraint of no partial or total overlap ({\it hard} dimers).
 The statistics of this model is described by the configurational
macrocanonical partition function
\begin{equation}
\Xi_N(z)=1+  \sum_{s=1}^{N/2} g_N(s)z^s.
\label{mdmpfN}
\end{equation}

Here $g_N(s)$ denotes the number of ways of placing $s$ dimers over
  the links of the
 $N$-site lattice, and $z$ is the dimer activity.

In the thermodynamical limit  

\begin{equation}
\Xi(z)=\lim_{N\rightarrow \infty}[\Xi_N(z)]^{1/N} 
=1+  \sum_{s=1}^{\infty} g(s)z^s
\label{mdmpf}
\end{equation}

and the macrocanonical potential  is defined as

\begin{equation}
\Gamma(z)={\rm ln}(\Xi(z)) 
= \sum_{s=1}^{\infty}\gamma_sz^s.
\label{mpot}
\end{equation}

The dimer density per site $\rho(z)$ is then expressed in terms of the
 potential $\Gamma(z)$ as

\begin{equation}
\rho(z)=z \frac{d\Gamma} {dz}= \sum_{s=1}^{\infty} d_n z^n  
\label{dimden}
\end{equation}
 with $d_n=n\gamma_n$.  Let us now summarize the simple
argument\cite{gaunt,baker,fisher,kurtze} relating the HT and low-field
expansions of the spin-1/2 Ising model in the limit of infinite
temperature with the low-density expansions for hard dimers on the
same lattice.  In the thermodynamical limit, the HT expansion of the
Ising magnetization per spin can be written as
\begin{equation}
M(H,T)=c +2(1-c^2)\sum_{l=1}^{\infty} \frac {v^l} {l} 
\sum_{m=1}^{l} m \psi_{ml}c^{2m-1}
\label{isimag}
\end{equation}
 with $v={\rm tanh}(K)$ and $c={\rm tanh}(h)$.  In the elementary graphical
representation of the HT and low-field expansion of the specific free
energy, each term of order $l$ in $v$ and $2m$ in $c$ is associated to
a configuration of $l$ bonds and $2m \le 2l $ marked vertices on the
lattice.  The coefficients surviving in the particular limit in which $v
\rightarrow 0$ and $c \rightarrow \infty $ with fixed $z=vc^2$, are
associated to configurations of $l$ separated bonds whose endpoints
are $2l$ marked vertices and therefore they are related to the number of
ways of placing $l$ hard dimers on the lattice in such a way that for
the magnetization (i.e. the field-derivative of the free energy) one has
\begin{equation}
M(H,T)/c \rightarrow [1-2\rho(z)]
\label{isirho}
\end{equation}
with
\begin{equation}
\rho(z)=\sum_{l=1}^{\infty} \psi_{ll} z^l
\label{rho}
\end{equation}
 the activity expansion of the dimer density.

 The  behavior eq.(\ref{magn}) of the magnetization nearby the YL edge
 must translate into
\begin{equation}
\rho(z) \sim |z-z_0|^{\sigma}
\label{rhosing}
\end{equation}
 for the dimer density, where 
\begin{equation}
z_0 =-\lim_{T\rightarrow \infty} ({\rm tan}(h''_{YL}(T))^2 {\rm tanh}(K)
\end{equation}
is a real negative quantity.
\section{ The extended  expansions of the dimer density}
  Our recent calculation\cite{bp1,bp2}, in most cases through order
 24, of the HT expansion of the spin-1/2 Ising magnetization in the
 presence of an external field, for several bipartite lattices of
 dimensionality $2 \le d \le 7$, enables us to extend through the
 same order the expansions of the dimer density in powers of the
 activity $z$ on these lattices.  Such expansions were already
 tabulated in the literature\cite{kurtze} in the case of the
 square(sq)(17 terms), the simple-cubic(sc)(15 terms), the
 body-centered-cubic(bcc)(13 terms) lattices. The known coefficients are
 reproduced by our calculation. Actually, a few more coefficients
 could be derived from the published HT and low-field
 expansions\cite{McK} for the sq(19 terms) and the sc(17 terms)
 lattices. Also for the hyper-simple-cubic(h4sc) lattice in $d=4$
 spatial dimensions, 16 correct terms of the free-energy expansion
 were tabulated\cite{McK} in the same paper.  However, these data were
 overlooked in Ref.[\onlinecite{kurtze}] and in the successive work.

 We have extended the activity expansions in the case of the h4sc and
 the hyper-bcc (h4bcc) lattices in $d=4$ spatial dimensions through
 the 24th order.  Our calculation for the higher-dimensional hyper-bcc
 lattices extends also through the 24th order, whereas in the case of the
 five- and six-dimensional hsc lattices, it reaches only the 22nd and the
 21st, respectively.  For dimension $d=7$, our expansion on the
 hsc lattice extend only to the 20th order.  These
 data were not yet available in the literature.

The expansions for the h5sc, h6sc,...,h7sc lattices have been
 truncated at a slightly smaller order, because on the (hyper-)cubic
 lattices the computation of the embedding numbers for the
 highest-order graphs is too time-consuming to be completed in just a
 few days by a single personal computer.  While in the case of the
 hbcc lattices the computational complexity of the calculation is
 independent\cite{bp1} of the lattice dimensionality, in the hsc case
 it grows exponentially with the space dimensionality.

  We have not yet extended the calculation of the series in the cases
 of non-bipartite lattices such as the triangular, face-centered-cubic
 and their high-dimensional generalizations.  The coefficients of our
 extended expansions of the dimer density series for $2 \le d\le 7$
 are reported in the Tables \ref{tab1}... \ref{tab5}.

\section{Analysis of the dimer expansions}
 For the dimer density series, we 
have located the nearest negative singularity in the complex
 $z$-plane  and estimated its exponent
 $\sigma$ by the numerical tools currently used to extrapolate the
 behavior of power expansions to the border of their analyticity disk,
 in particular employing the {\it differential approximants} (DAs).
 These are a generalization of the well known Pad\'e approximant(PA)
 method that approximates the series under study by the solution of an
 ordinary inhomogeneous linear differential equation with polynomial
 coefficients. For a detailed illustration of their definitions, of
 their properties, and for a general discussion of the uncertainties
 in the estimates of the singularity parameters determined by this
 method, we address the reader to
 Refs.[\onlinecite{gutlib,bp1,bp2,gutt}].

\begin{figure}[tbp]
\begin{center}
\leavevmode
\includegraphics[width=3.37 in]{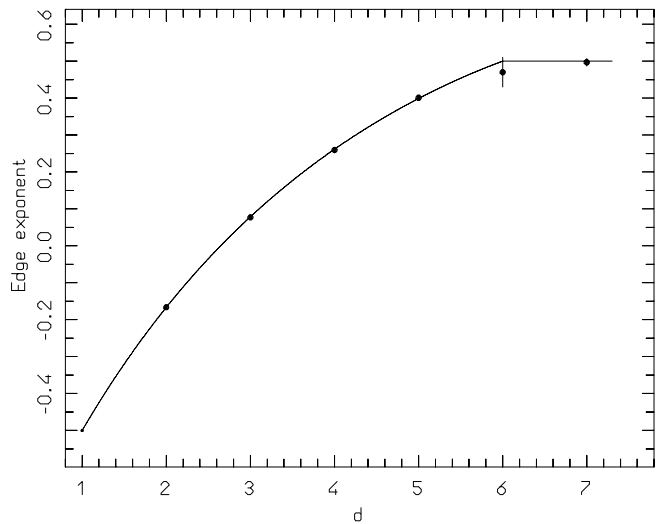}
\caption{ \label{figura_edge} Estimates (full circles)
of the edge-exponent $\sigma(d)$, obtained our dimer density expansions
 vs the lattice dimensionality. The continuous curve is obtained 
resumming the third order $\epsilon$-expansion of $\sigma(d)$    by a [3/2]
 Pad\'e approximant constrained to reproduce the exactly known values of
 the exponent for $d=1$ and $d=2$.  }
\end{center}
\end{figure}

Let us first consider the  few known terms of 
 the $\epsilon$-expansion of the exponent $\sigma$
\begin{equation}
\sigma(d)=\frac{1} {2} - 
\frac {1} {12} \epsilon-  \frac {79} {3888}  \epsilon^2
 + (\frac {1} {81} \zeta(3) -\frac {10445} {1250712})\epsilon^3 +O(\epsilon^4) 
\label{sigmaeps}
\end{equation}
 where $\zeta(n)$ denotes the Riemann zeta function.  The expansion
 eq.(\ref{sigmaeps}) can be resummed following various procedures, all
 yielding consistent results for $1<d<6$. For example, several simple
 continuous interpolations of $\sigma(d)$ with respect to the space
 dimensionality $d$, can be found\cite{kurtze,lai} in the literature,
 which are obtained from low-order constrained Pad\'e approximants.
 The continuous curve in Fig.\ref{figura_edge}, is the plot vs $d$ of
 a [3/2] Pad\'e approximant formed from the $\epsilon$-expansion
 eq.(\ref{sigmaeps}), that satisfies the additional constraints of
 reproducing the known exact values\cite{cardy} $\sigma(2)=-1/6$ for
 $d=2$ and $\sigma(1)=-1/2$ at the lower critical dimension $d=1$.
 For $d>6$, the exponent should retain the value $\sigma(d)=1/2$ and
 so this part of the curve is drawn flat.  The constrained [3/2] PA
 has the expression
\begin{equation}
\sigma(d)=\frac {-0.00564924115 \epsilon^3  -0.141867697 \epsilon^2 
+ 0.390189052 \epsilon+1/2} {-0.0852567479 \epsilon^2 +
 0.947044730 \epsilon + 1}.
\label{sigma32pa}
\end{equation}
The resummation by this approximant is visually indistinguishable from
 that obtained by the [6/0] PA subject to the constraint of
 reproducing the exact results for $d=0,1,2$. The latter approximation
 was proposed in Ref.[\onlinecite{lai}] to describe the behavior of
 the exponents of the unphysical singularity at negative activity for
 repulsive-core fluids as $d$ varies.  Alternative, but completely
 consistent interpolations can be obtained by other more refined
 resummation procedures\cite{bon}, including a Pad\'e-Borel method and
 a conformal mapping technique.  For each dimension $d$, the small
 spread among the estimates of $\sigma(d)$ obtained by the various
 resummations methods of the expansion eq.(\ref{sigmaeps}) can be used
 to roughly limit an interval of possible values of $\sigma(d)$, whose
 width increases with $\epsilon$. We have reported these spreads in
 the eighth column of Table \ref{tab6}.

 An $\epsilon$-expansion through the third order has been
 computed\cite{lai} also for the leading correction-to-scaling
 exponent $\theta(d)$. Its values $\theta(2)=0.83(1)$, 
$\theta(3)=0.622(12)$,
 $\theta(4)=0.412(8)$ and $\theta(5)=0.205(5)$ for the various
 dimensions have been estimated\cite{lai} by constrained PAs.  
 Here we shall  use these estimates without further discussion.

 From these studies of the $\epsilon$-expansion, we are thus led to
 expect that the exponent $\sigma(d)$ should lie between
 $\sigma=-1/2$, its $d=1$ value, and $\sigma=1/2$, the $d=6$
 value. For intermediate values of $d$, the exponent should take
 values not far from those suggested by the various resummation
 procedures.  In conclusion, $\sigma(d)$ is expected to take small and
 possibly positive values for $2< d < 6$.  The DAs are probably unable
 to measure accurately exponents in this range of values and
 therefore, as suggested in Ref.[\onlinecite{kurtze}], it is more
 convenient to analyze the $z$-derivative of the dimer density, rather
 than $\rho(z)$ itself, because  it has a sharper singularity,
 characterized by the exponent $ \sigma-1$.
\begin{table}[ht]\scriptsize
\caption{ Estimates of the location $z_0$ and the exponent $\sigma(d)$
 of the nearest singularity of the dimer density in the complex
 activity plane $z$ for several bipartite lattices of dimension $2 \le
 d \le 7$.  In the first column, we have indicated the structure and
 dimensionality of the lattice.  The second and fourth column show the
 results of Ref.[\protect\onlinecite{kurtze}], while the third and fifth
 contain the results of this paper. Notice that, for $ d\ge 5$, our
 estimates for the hsc lattices are based on shorter series.  In the
 sixth column, we have reported the values of $\sigma(d)=\phi(d)-1$
 obtained\protect\cite{lai} from the exponent occurring for repulsive-core
 fluids.  In the seventh column, we have shown estimates\protect\cite{hsu} of
 the exponent $\sigma(d)=\phi_I(d+2)-2$ derived from a high-precision
 MonteCarlo simulation of site-animals. In the eighth column, we have
 indicated ranges of estimates of $\sigma$ obtained from different
 resummation prescriptions of the $\epsilon$-expansion. The last
 column reports exact results for $\sigma$,
  whenever known. The data in the last
 two columns depend only on the lattice dimensionality.  }
\begin{tabular}{|c|c|c|c|c|c|c|c|c|}
 \hline
lattice  & $z_0$ (Ref.[\onlinecite{kurtze}]) & $z_0$ (Our work)&
 $\sigma$ (Ref.[\onlinecite{kurtze}])& $\sigma$ (Our work)
&Fluids (Ref.[\onlinecite{lai}])&Animals (Ref.[\onlinecite{hsu}])
& $\sigma$ $\epsilon$-exp.(Ref.[\onlinecite{bon}])& $\sigma$ Exact   \\
 \hline
${\rm sq}$&-0.088962(1)&-0.0889642(2)&-0.1645(20)
& -0.1662(5)&-0.161(8)&-0.165(6)&0.146-0.166
&-1/6    \\
${\rm tri}$&-0.056076(2)&&-0.1620(15)&&&&&\\
${\rm sc}$&-0.052002(6) &-0.0520268(2)&0.096(6)& 0.077(2)&0.0877(25)&0.080(7)
 &0.079-0.091&\\
${\rm bcc}$ &-0.037309(4) & -0.0373125(2) &0.090(5)& 0.076(2)&&&& \\
${\rm fcc}$ &-0.02413(18)&& 0.097(11)&&&&&\\
${\rm h4sc}$ & &-0.0365624(4)&  & 0.258(5)&0.2648(15)
&0.261(12)&0.262-0.266 &     \\
${\rm h4bcc}$ & &-0.0170399(2)&  &0.261(4) &&&&    \\
${\rm h5sc}$ & & -0.0281849(4)  &&0.401(9)&0.402(5)&0.40(2)&0.399-0.400 & \\
${\rm h5bcc}$ & & -0.00814235(4) &     &0.402(2)&&&  &     \\
${\rm h6sc}$ &  & -0.0229462(6) & &0.460(50)    &0.465(35)&&1/2&1/2 \\
${\rm h6bcc}$ &  & -0.0039832(2)& &0.475(30)    &&&& \\
${\rm h7sc}$ &  & -0.019360(3)& &0.495(8)    && &&1/2 \\
${\rm h7bcc}$ &  & -0.0019715(3)& &0.498(3)    && && \\
 \hline
\end{tabular} 
\label{tab6}
\end{table}

 Our procedure of numerical analysis is the following. To begin with,
we determine a central value $z_0$ of the location of the singularity
by averaging over an appropriate sample of estimates from first- or
second-order DAs, selected among those using most or all available
series coefficients. Evident outliers are excluded from the sample.
The uncertainty attached to $z_0$ is a small multiple of the spread of
the reduced sample.  Then we can evaluate also the corresponding
exponent $\sigma(d)$ using a sample of second- or third-order DAs
which are {\it biased} with the previously determined central value
$z_0$, i.e. are built to be singular at $z_0$.  This computational
procedure allows only partially for the influence of the expected
corrections to scaling on the estimates of the exponent, just because
the DAs are more flexible than PAs.  However the estimates so
obtained still suggest a significant residual influence of the
corrections to scaling, which is especially relevant in the case of
the (hyper)-sc lattices and for large dimensionalities, and so we have tried
to further improve our accuracy using also the well known
procedure\cite{rosk,lai} of evaluating the exponent after performing
the variable transformation
\begin{equation}
w=1-(1-z/z_0)^{\theta(d)}
\label{rosk}
\end{equation}
on the series. By this simple prescription, the leading non-analytic
correction to scaling of the given series is approximately changed
into an analytic one, which should be more accurately dealt with by
the DAs. We observe that generally the estimates show little
sensitivity to small variations of the bias value of $z_0$ and of the
correction exponents $\theta(d)$, so that our final uncertainties can
generously allow also for the spreads of these input parameters.

The estimates of the singularity location $z_0$ in the activity plane
and of the exponent $\sigma(d)$ obtained by this kind of analysis are
shown in the third and fifth columns of Table \ref{tab6} and compared
with those from a previous accurate DA analysis\cite{kurtze} of
activity expansions shorter than those now available, that was
performed along similar lines, but not explicitly allowing for the
corrections to scaling.  Our results generally lie within the range
roughly suggested by the resummed $\epsilon$-expansion and are
essentially consistent with the existing pure DA
estimates\cite{kurtze,gaunt} for $d<4$, but are more precise.  It is
worth to stress that our expansions of the dimer-density cover the
whole range of dimensionalities between the lower and the upper
critical dimension and therefore our results are also more
extensive. For comparison, in the Table \ref{tab6} for each value of
$d$, we have also reported estimates of $\sigma(d)$ obtained\cite{lai}
from the exponents $\phi(d)$ associated to repulsive-core fluids and
estimates obtained from the exponents $\phi_I(d+2)$ associated to the
undirected site animals in $d+2$ dimensions. The former data were
obtained\cite{lai} by a DA analysis (which allowed for the corrections
to scaling) of the 14th-order fugacity-expansions for a model of a
binary molecular mixture, with Gaussian interaction, in dimensions $1
\le d \le 6$.  The latter data were more recently obtained\cite{hsu}
by extensive MonteCarlo simulations performed in the case of
(hyper)-simple-cubic lattices with $2 \le d \le 9$.  Both sets of
estimates are essentially consistent with our results within their
(sometimes larger) uncertainties.

 It is also worth noticing that, if we allow explicitly for the
corrections to scaling using the variable transformation defined by
eq.(\ref{rosk}), our estimates of $\sigma(3)$ are somewhat smaller
than the older results\cite{kurtze} from shorter dimer series, show a
closer agreement with the constrained PA resummation of the
$\epsilon$-expansion and exhibit more accurately the expected
universality with respect to the lattice structure. The universality
check is satisfactory also in $d=4$. For $d=5$ dimensions, the check
is still as accurate, in spite of the small leading correction
exponent, because of which other unaccounted higher corrections to
scaling might also be significant.

   At the upper critical dimension $d_c=6$, it is expected that the
power singularity of the dimer density is multiplicatively
corrected\cite{lai,ruiz,ferb} by a power of a logarithm and that
moreover also the leading corrections to scaling are logarithmic.
Since the DAs are notoriously unfit to describe accurately this kind
of singularity, it is not surprising that their convergence
deteriorates. For lack of established and effective prescriptions to
deal with this computation, we cannot do better than assigning to the
estimate of $\sigma(6)$, an uncertainty much larger than the apparent
spread of DAs.

For completeness, we have computed $\sigma(d)$ also for the
higher-dimensional hsc and hbcc lattices just to check that, as
expected, above the upper critical dimension, the exponents
$\sigma(d)$ retain the mean-field value, so that $\sigma(7)=1/2$,
etc..  For graphical convenience, we have reported in the figure only
the seven-dimensional result.

In the Fig.\ref{figura_edge}, we have always reported the weighted
average of the estimates for the sc- and the bcc-type lattices and
attached to it a small multiple of the largest uncertainty of the
single results.

\section{Dimer constants}
 Let us now show briefly that sufficiently long low-activity expansions of
 the dimer density can be of use also in estimating heuristically the
 constant $h_d$ that controls the exponential growth of the number of
 all possible dimer arrangements over the bonds of a $N$-site
 $d$-dimensional simple-cubic lattice in the large $N$ limit. This
 quantity is defined by
\begin{equation}
 h_d= {\rm ln}[\lim_{N \rightarrow \infty}(\Xi_N(z))^{1/N}]_{z=1}=\Gamma(1)
\end{equation}
and is often  called {\it monomer-dimer constant} of the lattice under study, 
 the term {\it monomer} referring to a site unoccupied by a dimer.  

It is more difficult to get from the series data good estimates also
 for the constant that characterizes the exponential growth of the
 number of the closely-packed dimer coverings (i.e. dimer arrangements
 such that every lattice site belongs to a dimer) for a $N$-site
 $d$-dimensional simple-cubic lattice in the large $N$ limit,
\begin{equation}
\tilde h_d=\lim_{z\rightarrow \infty}[\Gamma(z) -\rho(z) {\rm ln} z].
\label{tilde_h}
\end{equation}
The quantity $e^{ \tilde h_d}$ is often called {\it molecular
freedom}. 

 The constants\cite{finch} $h_d$ and $\tilde h_d$ (which of course
have nothing to do with the reduced magnetic field mentioned in the
Introduction), are of interest in chemistry, combinatorial mathematics
and information theory, but unfortunately their exact values are not
known except\cite{Kast,FisTe,Fisdim} in the case of $\tilde h_2$. In
particular, the quantities $h_2 $, $h_3 $, $h_4$ and $h_5 $ are known
rigorously only through their bounds.

Consider first the computation of $ h_d$.  In the case of the sq
 lattice, a naive [12,12] PA of the $z$-expansion of $\Gamma(z)$
 yields the estimate $h_2 \approx 0.662798...$. This result agrees to
 six significant digits with the most precise
 determination\cite{baxter,fried1} $h_2 = 0.6627989727(1)$ and is
 consistent with the tightest presently known bounds $0.66279897190
 \le h_2 \le 0.662798972844913$. Other accurate, but also non-rigorous
 estimates of $h_2$ can be found in the
 literature\cite{gaunt,runnels,nagle,beichl,kong,huo}. In particular
 Ref.[\onlinecite{kong}] reports a twelve figure determination of this
 constant.

By the same quite simple prescription, in the case of the sc lattice,
 we find $h_3 \approx 0.7859...$, a value which lies between the
 known\cite{fried2} bounds: $0.7849602275 \le h_3 \le 0.7862023450$.
 For the h4sc lattice, our PA estimate is $h_4 \approx 0.880...$,
 which should be compared with the lower bound\cite{fried1}
 $0.8638570485$.  These results show that very naive PAs yield rather
 accurate analytic continuations of the $z$-expansion of $\Gamma(z)$
 well outside its very small convergence disk. The precision of the
 estimates decreases as the space dimension $d$ grows simply because
 the radius of convergence $|z_0|$ of the $z$-expansion of $\Gamma(z)$
 vanishes as $d$ grows (see Table \ref{tab6}).

 Unfortunately, this straightforward approach does not provide an
 analytic continuation of $\Gamma(z)$ of acceptable accuracy for
 $z>>1$, that is necessary to compute also $\tilde h_d$. We can try to
 mitigate this difficulty, performing in the $z$-expansion the simple
 change\cite{nagle,heil} of variable $z=t/(1-(q-1)t)^2$, where $q$ is
 the coordination number of the lattice under study. As a first
 result, we can observe a drastic improvement in the apparent accuracy
 of the aforementioned estimates of the constants $h_d$. For example,
 we obtain $h_2 \approx 0.662798972..$, which reproduces three
 additional digits of the best estimate of $h_2$. For the
 three-dimensional sc lattice, the improved estimate is $h_3 \approx
 0.7859660...$ and in four dimensions we obtain $h_4 \approx
 0.88071788...$.  For the five-dimensional cubic lattice: $h_5 \approx
 0.958123...$, consistently with the known lower bound $ 0.94383303
 \le h_5$. Let us also stress that for the five- (and
 higher-dimensional) hsc lattices our expansions are slightly
 shorter. For example, they extend only through the 22nd order in the
 five-dimensional hsc case.
 
 Correspondingly, we have computed the values of the dimer density per
 site for $z=1$ obtaining: $\rho_2(1) \approx 0.31906155... $ (in
 complete agreement with the estimates of
 Ref.[\onlinecite{baxter,runnels}]), $\rho_3(1) \approx
 0.3421901... $, $\rho_4(1) \approx 0.3579234... $ and $\rho_5(1)
 \approx 0.369580...$.  Other estimates for higher-dimensional hsc
 lattices and for the hbcc lattices can also be found in Table
 \ref{tab7}. For $d>3$, our estimates of $h_d$ for
 the hsc lattices compare well with the
 lower bounds provided by the inequality\cite{fried1,fried2}
\begin{equation}
 h_d \geq  \frac{1}{2}[-p{\rm ln}p-2(1-p){\rm ln}(1-p)+p{\rm ln}2d -p]
\label{ineq}
\end{equation}
 with $p=\frac{4d +1-\sqrt{8d+1}}{4d}$.  The meaning of the estimates
 of $h_d$, summarized in Table \ref{tab7}, is the following: we have
 reported only the decimal figures which appear to be stabilized in
 the three highest-order diagonal PAs that can be formed from the
 available dimer expansions.

\begin{table}[ht]\scriptsize
\caption{ Our estimates of the dimer constants $h_d$ and the
 corresponding estimates of the dimer densities $\rho_d(1)$(third
 column) for (hyper-)simple-cubic and (hyper-)body-centered-cubic
  lattices of various dimensionalities. In
 the second column we have reported lower bounds and in the fourth
 upper bounds for $h_d$ on the hsc lattice when available. 
No bounds are known for the hbcc lattices. In the text we have reported also
 estimates by other authors, when existing.}
\begin{tabular}{|c|c|c|c|c|c|}
 \hline
 hsc & Lower limit& This work& Upper limit &hbcc& This work   \\
 \hline
  $h_2$  &0.66279897190 & 0.662798972(1)&0.662798972844913&& \\
 $\rho_2(1)$ & &0.31906155(1)& & &\\
  $h_3$  &0.7849602275 & 0.7859660(1)&0.7862023450& $h_3$ & 0.8813479(1)\\
 $\rho_3(1)$ & &0.3421901(1)& &$\rho_3(1)$&0.3584362(1)\\
  $h_4$  &0.8638570485 &0.880718(1)& & $h_4$ &1.131880(1)\\
 $\rho_4(1)$ & &0.3579234(1)& &$\rho_4(1)$&0.392552(1)\\
   $h_5$  &0.94383303 &0.95813(1)& &  $h_5$ & 1.408080(1)\\
 $\rho_5(1)$ & & 0.369580(1)&&$\rho_5(1)$&0.420248(1)\\
   $h_6$  & 1.01129436 & 1.023732(1)& &  $h_6$ & 1.703890(2)\\
 $\rho_6(1)$ &&0.37868106(1)& &$\rho_6(1)$&0.4417731(1)\\
  $h_7$ &1.06972606 &1.080759(2)&&    $h_7$&2.0142(2)\\ 
 $\rho_7(1)$ &&0.38604998(2)& & $\rho_7(1)$& 0.457881(2)\\
 \hline
 \end{tabular} 
 \label{tab7}
\end{table}

 In terms of the new variable, the potential $\Gamma(z)$ can be, with
 a better approximation, continued by PAs into a larger region of the
 complex $z$-plane, encompassing intermediate values of $z$ such as $z
 \approx 10$. We have tried to infer the large $z$ limit of the
 quantity $\Gamma(z)-\rho(z){\rm ln}z$ in eq.(\ref{tilde_h}) by
 observing that, in the intermediate $z$ region, this quantity behaves
 linearly in $1/\sqrt z$ to a fair approximation, and extrapolating
 this behavior to $z=\infty$.  We can thus conjecture estimates of the
 constants $\tilde h_d$, that although generally reasonable and
 consistent with the known\cite{fried1} bounds, unsurprisingly show an
 accuracy much lower than in the $h_d$ case and therefore will not be
 reported in the Table \ref{tab7}.  The blame for this failure lays
 on the still insufficient extension of the region of accurate
 analytic continuation and the arbitrariness of the various
 extrapolation procedures, that make a safe evaluation of the
 uncertainties difficult.
 
\section{Conclusions}
 We have derived the coefficients of the series expansions of the
dimer density in powers of the activity, in most cases through 24th-order,
for several bipartite lattices of dimensionality $2\le d \le 7$. Thus
we have not only extended the existing series data, so far published
only for $d \le 3$, but also produced series for $d \ge 4$.  An
analysis of this set of data improves the accuracy in the
determination of the locations and the exponents of the nearest real
negative singularity of the dimer density in the complex activity
plane.  The exponents controlling the Yang-Lee edge-singularity of
this class of ferromagnetic spin models in a wide range of space
dimensions are related, in some cases by a simple dimensionality
shift, with the exponents characterizing very different systems. The
numerical estimates presently available for all these exponents show a
good consistency among them and with the results of appropriate
resummations of the renormalization-group $\epsilon$-expansion. Our
dimer-density series are also shown to be of some use in estimating
the dimer constants $ h_d$ and $\tilde h_d$.

\section{Acknowledgements}
We are grateful to A.J. Guttmann for reading a preliminary draft. We
 are also grateful to P.H.Lundow and K. Markstr\"om for helpful
 comments on the final Section of the paper.  We thank the Physics
 Departments of Milano-Bicocca University and of Milano University for
 hospitality and support.  Partial support by the MIUR is also
 acknowledged.

\end{document}